\pdfoutput=1 % pdfoutput=1 is for arXiv
\newif\ifarXiv
\arXivtrue % comment out to switch to journal template
\ifarXiv
\documentclass[aps,pre,reprint,superscriptaddress]{revtex4-2}

\else
\documentclass{rsproca_new}
\jname{rspa}
\Journal{Proc R Soc A\ }
\fi

\newcommand{\nn}{\nonumber}
\newcommand{\ie}{{i.e.},~}
\newcommand{\eg}{{e.g.},~}
\newcommand{\etal}{{et al.}~}
\newcommand{\fig}[1]{Figure~\ref{fig:#1}}

\newcommand{\secRef}[1]{Section~\ref{sec:#1}}
\newcommand{\appRef}[1]{Appendix~\ref{app:#1}}
\newcommand{\eq}[1]{Eq.~(\ref{eq:#1})}
\usepackage{amsmath,amssymb,bm,siunitx,mathtools}
\usepackage{cleveref} % referencing range of equations
\usepackage{graphicx}% Include figure files
\graphicspath{{images/}}
\usepackage{bm}% bold math
\usepackage{booktabs,multirow} % tables
\usepackage{blkarray} % block matrices with indices
\usepackage{lineno}

\begin{document}

\title{Deriving pairwise transfer entropy from network structure and motifs}

\ifarXiv
%\preprint{}


\author{Leonardo Novelli}
\email[]{leonardo.novelli@sydney.edu.au}
%\thanks{}
\affiliation{Centre for Complex Systems, Faculty of Engineering, The University of Sydney, Sydney, Australia}
\author{Fatihcan M. Atay}
%\thanks{}
\affiliation{Department of Mathematics, Bilkent University, 06800 Ankara, Turkey}
\affiliation{Max Planck Institute for Mathematics in the Sciences, Inselstra{\ss}e 22, 04103 Leipzig, Germany}
\author{J{\"u}rgen Jost}
%\thanks{}
\affiliation{Max Planck Institute for Mathematics in the Sciences, Inselstra{\ss}e 22, 04103 Leipzig, Germany}
\affiliation{Santa Fe Institute for the Sciences of Complexity, Santa Fe, New Mexico 87501, USA}
\author{Joseph T. Lizier}
%\thanks{}
\affiliation{Centre for Complex Systems, Faculty of Engineering, The University of Sydney, Sydney, Australia}
\affiliation{Max Planck Institute for Mathematics in the Sciences, Inselstra{\ss}e 22, 04103 Leipzig, Germany}


\date{\today}

\else

\author{%%%% Author details
Leonardo Novelli$^{1}$, Fatihcan M. Atay$^{2,3}$, J{\"u}rgen Jost$^{3,4}$ and Joseph T. Lizier$^{1,3}$}

%%%%%%%%% Insert author address here
\address{$^{1}$Centre for Complex Systems, Faculty of Engineering, The University of Sydney, Sydney, Australia\\
$^{2}$Department of Mathematics, Bilkent University, 06800 Ankara, Turkey\\
$^{3}$Max Planck Institute for Mathematics in the Sciences, Inselstra{\ss}e 22, 04103 Leipzig, Germany\\
$^{4}$Santa Fe Institute for the Sciences of Complexity, Santa Fe, New Mexico 87501, USA}

%%%% Subject entries to be placed here %%%%
\subject{Applied Mathematics, Network Neuroscience}

%%%% Keyword entries to be placed here %%%%

%%%% Insert corresponding author and its email address}
\corres{Leonardo Novelli\\
\email{leonardo.novelli@sydney.edu.au}}

\fi

\begin{abstract}
% motivation
Transfer entropy is an established method for quantifying directed statistical dependencies in neuroimaging and complex systems datasets. 
% problem/question
The pairwise (or bivariate) transfer entropy from a source to a target node in a network does not depend solely on the local source-target link weight, but on the wider network structure that the link is embedded in.
% method
This relationship is studied using a discrete-time linearly-coupled Gaussian model, which allows us to derive the transfer entropy for each link from the network topology.
% results
It is shown analytically that the dependence on the directed link weight is only a first approximation, valid for weak coupling.
More generally, the transfer entropy increases with the in-degree of the source and decreases with the in-degree of the target, indicating an asymmetry of information transfer between hubs and low-degree nodes. 
In addition, the transfer entropy is directly proportional to weighted motif counts involving common parents or multiple walks from the source to the target, which are more abundant in networks with a high clustering coefficient than in random networks.
% significance
Our findings also apply to Granger causality, which is equivalent to transfer entropy for Gaussian variables.
Moreover, similar empirical results on random Boolean networks suggest that the dependence of the transfer entropy on the in-degree extends to nonlinear dynamics.
\end{abstract}

\maketitle

%\linenumbers

\section{Introduction}
From a network dynamics perspective, the activity of a system over time is the result of the interplay between the dynamical rules governing the nodes and the network structure (or topology).
Studying the structure-dynamics relationship is an ongoing research effort, often aimed at optimising the synchronisation, controllability, or stability of complex systems, or understanding how these properties are shaped by evolution~\cite{Barrat2008,Liu2016,Nishikawa2017,Sporns2000}.
Information theory~\cite{Shannon1948} offers a general mathematical framework to study the diverse range of dynamics across technical and biological networks, from neural to genetic to cyber-physical systems~\cite{Prokopenko2009}. It provides quantitative definitions of uncertainty and elementary information processing operations (such as storage, transfer, and modification), which align with qualitative descriptions of dynamics on networks and could serve as a common language to interpret the activity of complex systems~\cite{Lizier2013}.

This study will focus on a specific information-theoretic measure: transfer entropy~(TE)~\cite{Schreiber2000,Bossomaier2016}. In its original formulation as a pairwise measure, TE can be used to study the activity of a network and detect asymmetric statistical dependencies between pairs of nodes.
TE has been widely used to characterise directed relationships in complex systems, in particular in the domain of computational neuroscience~\cite{Wibral2014,Timme2018}.
For a given dynamics, there is a non-trivial dependence of the local TE between pairs of nodes and the wider global structure of the network. 
For example, several empirical studies have reported a dependence of the TE on the in- and out-degree of the source and target nodes~\cite{Marinazzo2014,Marinazzo2012,Timme2016,Lizier2009,Ceguerra2011,Li2019} as well as other aspects of network structure such as long links in small world networks~\cite{Lizier2011}.
The main purpose of this work is to present a systematic analytic characterisation of the relationship between network structure and TE on a given link, which has not been previously established. 
% -- elucidating this relationship is the main purpose of this work.

In order to provide an analytic treatment, we will use a stationary vector autoregressive (VAR) process, characterised by linear interactions and driving Gaussian noise (\secRef{preliminary}). This model is a simplification as compared to most real-world processes, but can be viewed as approximating the weakly coupled near-linear regime~\cite{Barnett2009a}. Interestingly, a recent review found that the VAR model performed better than six more complex mainstream neuroscience models in predicting the \emph{undirected} functional connectivity (based on Pearson correlation) from the brain structural connectivity (based on tractography)~\cite{Messe2015}. 
Other studies have related the undirected functional connectivity to specific structural features, such as search information, path transitivity~\cite{Goni2014}, and topological similarity~\cite{Bettinardi2017}. Analytic relationships of the network structure and correlation/covariance between nodes for the VAR and similar dynamics have also been well studied~\cite{Galan2008, Pernice2011, Saggio2016}.

This work will instead focus on the analytical treatment of the \emph{directed} functional connectivity obtained via the pairwise TE for the VAR process.
Building on previous studies of other information-theoretic measures in this process (regarding the TSE complexity \cite{Tononi1994} in ~\cite{Barnett2009a,Barnett2011} and active information storage in \cite{Lizier2012}), we explicitly establish the dependence of the TE for a given link on the related structural motifs.
Motifs are small subnetwork configurations, such as feedforward or feedback loops, which have been studied as building blocks of complex networks~\cite{Milo2004}. Specific motif classes are over-represented in biological networks as compared to random networks, suggesting they could serve specific functions~\cite{Song2005,Sporns2004,Mangan2003,Azulay2016}. Indeed, linear systems analyses have been used to predict functional sub-circuits from the nervous system topology of the \textit{C.~elegans} nematode~\cite{Varshney2011}.

It is shown analytically (in \secRef{derivations}) that the dependence of the TE on the directed link weight from the source to the target is only a first approximation, valid for weak coupling.
More generally, the TE increases with the in-degree of the source and decreases with the in-degree of the target, indicating an asymmetry of information transfer between hubs and low-degree nodes. 
In addition, the TE is directly proportional to weighted motif counts involving common parents or multiple walks from the source to the target, which are more abundant in networks with a high clustering coefficient than in random networks. These results are tested using numerical simulations and discussed in \secRef{simulations}.

Being based on a linearly-coupled Gaussian model, our findings apply directly to Granger causality, which is equivalent to TE for Gaussian variables~\cite{Barnett2009b}. However, similar empirical results on random Boolean networks (RBNs) suggest that the dependence of the TE on the in-degrees extends to nonlinear dynamics~(\appRef{RBN}).

\section{Information-theoretic measures on networks of coupled Gaussians\label{sec:preliminary}}
Let us consider a discrete-time, stationary, first-order autoregressive process on a network of $N$ nodes. This multivariate VAR(\num{1}) process is described by the recurrence relation
\begin{equation}\label{eq:VAR}
    \bm{Z}(t+1)=\bm{Z}(t)\cdot C+\bm{\varepsilon}(t),
\end{equation}
where $Z_i(t)$ is the activity of node~$i$ at time~$t$ (and $\bm{Z(t)}$ is a row vector). Here, $\bm{\varepsilon}(t)$ is spatially and serially uncorrelated Gaussian noise of unit variance and $C=~[C_{ij}]$ is the $N\times N$ weighted adjacency matrix representing the weighted network structure (where $C_{ij}$ is the weight of the directed connection from node~$i$ to node~$j$). A stationary autoregressive process has a multivariate Gaussian distribution, whose expected Shannon entropy~\cite{Shannon1948}, independent of $t$, is~\cite[Ch. 8]{Cover2005}:
\begin{equation}\label{eq:entropy}
    H(\bm{Z})=\frac{1}{2} \ln[(2\pi e)^N |\Omega|].
\end{equation}
In \eq{entropy}, $|\Omega|$ represents the determinant of the covariance matrix $\Omega\coloneqq\langle\bm{Z}(t)^T \bm{Z}(t)\rangle$ and $\langle\cdot\rangle$ denotes the average over the statistical ensemble at times~$t$~\cite{Cover2005}. Barnett \etal\cite{Barnett2009a} show that the covariance matrix satisfies $\Omega=I+C^T\Omega C$, where $I$ denotes the relevant identity matrix, and the solution is obtained in general via the power series
\begin{align}\label{eq:omega_series}
    \Omega=I+C^T C+(C^2)^T C^2+\ldots=\sum_{j=0}^{\infty}(C^j)^T C^j.
\end{align}
(A simpler form exists for symmetric $C$~\cite{Barnett2009a}). As discussed in~\cite{Barnett2009a,Lizier2012}, the convergence of the series is guaranteed under the assumption of stationarity (for which a sufficient condition is that the spectral radius of $C$ is smaller than one).
Information-theoretic measures relating variables over a time difference~$s$ also involve covariances across time, which can be computed via the lagged covariance matrix~\cite{Lizier2012}
\begin{equation}\label{eq:omega_lag}
    \Omega(s)\coloneqq\langle\bm{Z}(t)^T \bm{Z}(t+s)\rangle=\Omega C^s.
\end{equation}
%where~$s$ is an integer time lag~\cite{Lizier2012}.
Interestingly, \eq{omega_lag} can be used to directly reconstruct the weighted adjacency matrix $C$ from empirical calculations of $\Omega$ and $\Omega(s)$ from observations~\cite{Lai2017}.

\section{Approximating the pairwise transfer entropy\label{sec:derivations}}
\begin{figure}
    \centering
    \ifarXiv\includegraphics[width=0.48\textwidth]{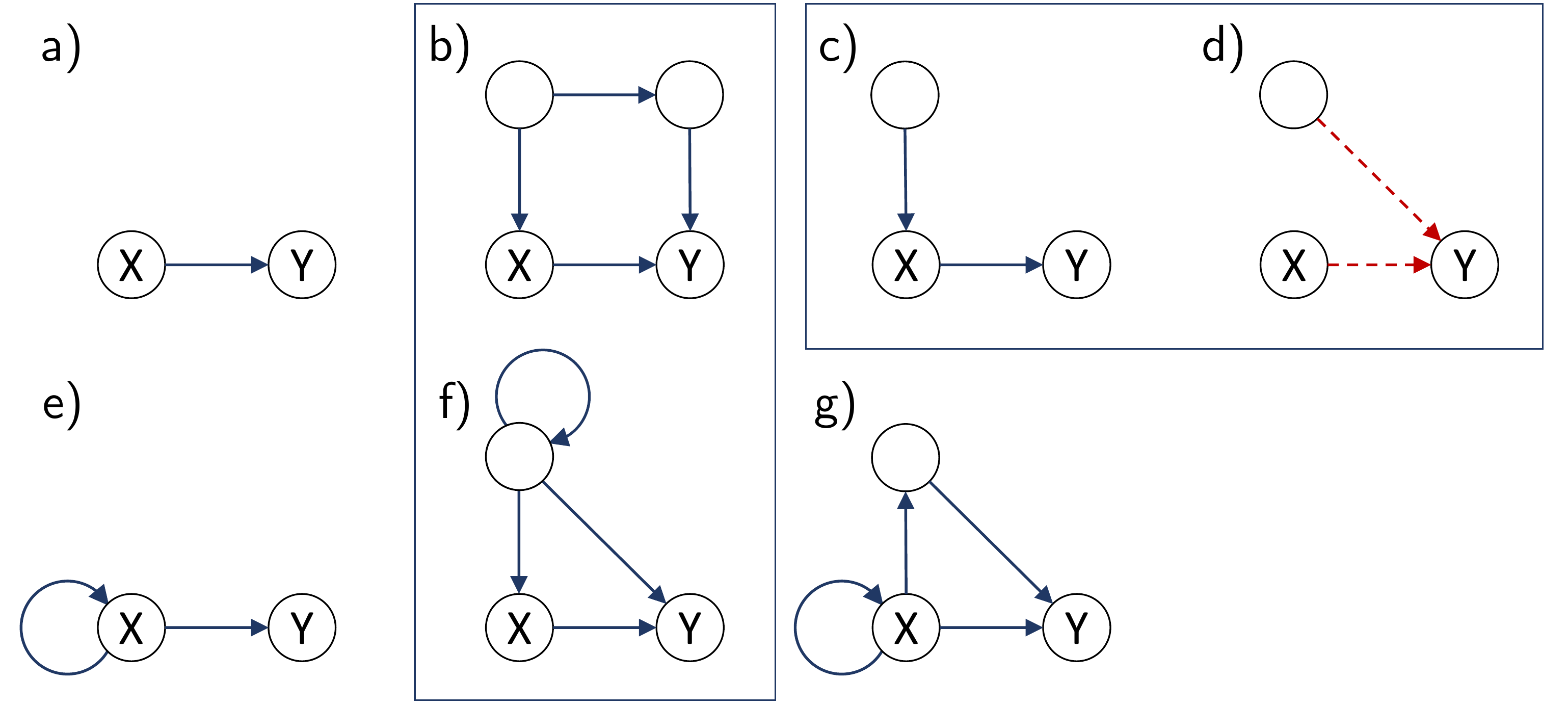}
    \else\centering\includegraphics[width=0.9\textwidth]{motifs_all}
    \fi
    \caption{\label{fig:motifs_all}
        Visual summary of the motifs involved in the pairwise transfer entropy from a source node~$X$ to a target node~$Y$ in the network. The seven panels (a-g) correspond to the seven motifs in \Crefrange{eq:motifs_all_a}{eq:motifs_all_g}, expanded up to order $\mathcal{O}(\|C\|^4)$. The motifs in panels~c and~d represent the effect of the weighted in-degree of the source and the target (which have a positive and negative contribution to the transfer entropy, respectively, with the negative indicated in dashed red line).
        The motifs in panels~b, f, and~g are clustered motifs, which can enhance or detract from the predictive effect of the directed link, depending on the sign of the link weights. In particular, motifs~b and~f involve a common parent of $X$ and $Y$, whereas~g involves an additional pathway effect. Note that the unlabelled nodes are distinct from $X$ and $Y$ (and from each other in panel b).
    }
\end{figure}

In this section, we will derive the TE~\cite{Schreiber2000} for pairs of nodes from the VAR process in~\eq{VAR} as a function of specific network motifs; the final results are listed in~\eq{motifs_all} and shown in~\fig{motifs_all}.

For two given nodes $X$ and $Y$ in $\bm{Z}$, the transfer entropy $T_{X\rightarrow Y}$ as a conditional mutual information can be decomposed into four joint entropy terms~\cite{Bossomaier2016}:
\begin{align}\label{eq:TE_entropies}
    T_{X\rightarrow Y}  = I(X;Y|\bm{Y^-}) = &H(Y,\bm{Y^-})-H(\bm{Y^-}) \nn \\
                        &-H(X,Y,\bm{Y^-})+H(X,\bm{Y^-}).
\end{align}
Here we use the shorthand $Y$ to represent the next value $Y(t+1)$ of the target at time $t+1$, $X$ for the previous value $X(t)$ of the source, and $\bm{Y^-}$ for the past \textit{state} of~$Y$ at time $t$. We drop the time index $t$ to simplify the notation under the stationarity assumption. Following convention, finite embedding vectors~$\bm{Y^-}\coloneqq\bm{Y^{(k)}}$ of the past $k$ values of $Y$ will be used to represent the previous state \cite{Schreiber2000,Bossomaier2016}. (One could also embed the source process $X$; however, only a single value is used here, in line with the order-\num{1} causal contributions in \eq{VAR}).

We can then rewrite the TE in terms of $\Omega(Y,\bm{Y^{(k)}}),\ \Omega(\bm{Y^{(k)}}),\ \Omega(X,Y,\bm{Y^{(k)}}),\text{ and }\Omega(X,\bm{Y^{(k)}})$: the covariance matrices of the joint processes involved in the four entropy terms.
Plugging \eq{entropy} into \eq{TE_entropies} for each term yields
\begin{align}\label{eq:TE_logs}
    T_{X\rightarrow Y}=\frac{1}{2} (&\ln|\Omega(Y,\bm{Y^{(k)}})|-\ln|\Omega(\bm{Y^{(k)}})| \nn\\
                        &-\ln|\Omega(X,Y,\bm{Y^{(k)}})|+\ln|\Omega(X,\bm{Y^{(k)}})|).
\end{align}
Furthermore, from the matrix identity $|e^A|=e^{\textnormal{tr}(A)}$ (valid for any square matrix $A$~\cite{Hall2015}) and from the Taylor-series expansion for the natural logarithm, it follows that
\begin{align}\label{eq:log_det}
    \ln|\Omega|=\displaystyle\sum_{m=1}^{\infty} \frac{(-1)^{m-1}}{m} \textnormal{tr}[(\Omega -I)^m],
\end{align}
where tr$[\cdot]$ is the trace operator. Plugging \eq{log_det} into \eq{TE_logs} gives
\begin{alignat}{2}\label{eq:TE_traces}
    T_{X\rightarrow Y}%&=\displaystyle\sum_{m=1}^{\infty}T_{X\rightarrow Y}^{\,m} \nn\\
    &=\frac{1}{2}\displaystyle\sum_{m=1}^{\infty} \frac{(-1)^{m-1}}{m}\big(&&\textnormal{tr}[(\Omega(Y,\bm{Y^{(k)}})-I)^m] \nn\\
    &{} &&-\textnormal{tr}[(\Omega(\bm{Y^{(k)}})-I)^m] \nn\\
    &{} &&-\textnormal{tr}[(\Omega(X,Y,\bm{Y^{(k)}})-I)^m] \nn\\
    &{} &&+\textnormal{tr}[(\Omega(X,\bm{Y^{(k)}})-I)^m]\big).
\end{alignat}
In order to simplify \eq{TE_traces}, consider the block structure of $B\coloneqq(\Omega(X,Y,\bm{Y^{(k)}})-I)$ and notice that it contains $(\Omega(Y,\bm{Y^{(k)}})-I),\ (\Omega(\bm{Y^{(k)}})-I),\text{ and }(\Omega(X,\bm{Y^{(k)}})-I)$ as submatrices with overlapping diagonals:
\begin{equation}\label{eq:B}
    B\coloneqq\Omega(X,Y,\bm{Y^{(k)}})-I=
    %https://tex.stackexchange.com/questions/10122/bordermatrix-with-blocks
    \begin{blockarray}{cccc}
            ~           & X      & Y      & \bm{Y^{(k)}} \\ 
        \begin{block}{c(c|c|c)}
            X           & \cdot  & \cdot  & \cdot   \\  
            \cline{2-4}
            Y           & \cdot  & \cdot  & \cdot   \\
            \cline{2-4}
            \bm{Y^{(k)}}\;\;   & \cdot  & \cdot  & \cdot   \\
        \end{block}
    \end{blockarray}\;\; - I =
\end{equation}
\begin{equation*}
    \left(
    \begin{array}{c|c|ccc}
        \Omega(0)_{XX}\!\!-\!\!1      & \Omega(1)_{XY} & \Omega(0)_{YX}  & \cdots & \Omega(k\!\!-\!\!1)_{YX} \\
        \hline
        \Omega(1)_{XY}   & \Omega(0)_{YY}\!\!-\!\!1    & \Omega(1)_{YY}    & \cdots & \Omega(k)_{YY} \\
        \hline
        \Omega(0)_{YX}      & \Omega(1)_{YY}      & \Omega(0)_{YY}\!\!-\!\!1  & \cdots & \Omega(k\!\!-\!\!1)_{YY} \\
        \vdots & \vdots     & \vdots  & \ddots      & \vdots \\
        \Omega(k\!\!-\!\!1)_{YX}        & \Omega(k)_{YY}      & \Omega(k\!\!-\!\!1)_{YY}  & \cdots & \Omega(0)_{YY}\!\!-\!\!1 \\
    \end{array}
    \right),
\end{equation*}
where $\Omega(s)_{XY}$ represents the $(X,Y)$ entry of the lag $s$ covariance matrix $\Omega(s)$ in \eq{omega_lag}.
An explicit representation of these covariance matrices is provided in \appRef{cov_matrices}. Since most of the terms in the trace of~$B^m$ also appear in the traces of the other covariance matrices in \eq{TE_traces}, they will get cancelled. As shown in \appRef{cov_matrices}, the only non-zero terms remaining in \eq{TE_traces} are those in $\textnormal{tr}[B^m]$ that involve multiplication of at least one entry of~$B$ from the first row or column (corresponding to correlations with $X$) \textbf{and} one entry from the second row or column (corresponding to correlations with the next value of the target $Y$). Therefore, we can simplify \eq{TE_traces} as
\begin{equation}\label{eq:TE_B}
    T_{X\rightarrow Y}=\frac{1}{2}\displaystyle\sum_{m=1}^{\infty}T_{X\rightarrow Y}^{\,(m)}=\frac{1}{2}\displaystyle\sum_{m=1}^{\infty}\frac{(-1)^{m}}{m}\overline{\textnormal{tr}[B^m]},
\end{equation}
where $T_{X\rightarrow Y}^{\,(m)}$ indicates contributions to $T_{X\rightarrow Y}$ from power $m$ of $B$, and the overbar on $\overline{\textnormal{tr}[B^m]}$ indicates that only the terms that involve at least one entry of~$B$ from the first row and one from the second row (or columns) are considered.
More formally,
\begin{align}
    \overline{\textnormal{tr}[B^m]}=&\overline{\sum_{i}(B^m)_{ii}} \\
    %=&\sum_{i_1,\ldots,i_{m}=1}^N B_{i_{1}}B_{i_{2}}\left(\prod_{j=2}^{m-1}B_{i_{j}i_{j+1}}\right)B_{i_{m}}B_{i_{1}} \nn\\
    =&\sum_{\substack{
        i_1,\ldots,i_{m} \text{ s.t.}\\
        \{1,2\}\subset \{i_1,\ldots,i_{m}\} \\
      }} B_{i_{1}i_{2}}B_{i_{2}i_{3}}\ldots B_{i_{m-1}i_{m}}B_{i_{m}i_{1}}. \nn
\end{align}

Let us now consider the cases $m=1,2$ separately. When $m=1$, all the terms in $\overline{\textnormal{tr}[B]}$ are neglected:
\begin{align}\label{eq:TE_m1}
    T_{X\rightarrow Y}^{\,(1)}=-\overline{\textnormal{tr}[B]}=-\overline{\sum_{i}B_{ii}}=0.
\end{align}
When $m=2$, we have
\begin{align}\label{eq:TE_m2}
    T_{X\rightarrow Y}^{\,(2)}=\frac{1}{2}\overline{\textnormal{tr}[B^2]}=&\frac{1}{2}\overline{\sum_{i,j}B_{ij}B_{ji}}=\frac{1}{2}\displaystyle\sum_{
        \substack{
           i=1;j=2 \\
           i=2;j=1
          }}B_{ij}B_{ji} \nn\\
      =&[\Omega(1)_{XY}]^2=[(\Omega C)_{XY}]^2,
\end{align}
where the last step follows from \eq{omega_lag}. Before proceeding to consider the cases $m>2$, let us see how \eq{TE_m2} can be used to relate the TE contribution $T_{X\rightarrow Y}^{\,(2)}$ to the network structure. Plugging \eq{omega_series} into \eq{TE_m2} yields
\begin{subequations}\label{eq:motifs_2}
    \begin{align}
        T_{X\rightarrow Y}^{\,(2)}  =&(C_{XY})^2+2C_{XY}(C^{T}C^2)_{XY}+\mathcal{O}(\|C\|^6) \nn\\
                                    =&(C_{XY})^2 \label{eq:motifs_2_1}\\
                                    &+2 \sum _{i_1,i_2} C_{XY}C_{i_1X}C_{i_1i_2}C_{i_2Y} \label{eq:motifs_2_2}\\
                                    &+\mathcal{O}(\|C\|^6). \nn
    \end{align}
\end{subequations}
In \eq{motifs_2} and in the following, we will only consider the contributions to the TE up to order $\mathcal{O}(\|C\|^4)$, where $\|\cdot\|$ is any consistent matrix norm~\cite{Barnett2009a}. Our approximations will therefore be most accurate when the link weights are homogeneous or have the same order of magnitude.
%longer past $k>2$ doesn't make any difference since the order of the corresponding terms is larger than~\num{4}
%\eq{motifs_2_1} is immediately interpretable as the \emph{directed effect} motif $X\rightarrow Y$. On the other hand, \eq{motifs_2_2} is more easily interpretable as a motif once we expand it as a sum (shown in \fig{motifs_2}):
%\begin{align}
%    2C_{XY}(C^{T}C^2)_{XY}=2 \sum _{i,j} C_{XY}C_{j,X}C_{j,i}C_{i,Y}.
%\end{align}
Noting that product sums of connected link weights as in \eq{motifs_2_2} represent weighted walk counts of relevant motifs,
the first two panels in~\fig{motifs_all} (panels a and b) provide a visual summary of the motifs involved in $T_{X\rightarrow Y}^{\,(2)}$.
%\begin{figure}
%    \includegraphics[width=0.3\textwidth]{motifs_2}
%    \caption{\label{fig:motifs_2}}
%\end{figure}

Now, consider the higher order cases. When $m=3$, we have
\begin{subequations}\begin{align}
    T_{X\rightarrow Y}^{\,(3)}=&-\frac{1}{3}\overline{\textnormal{tr}[B^3]}=-\frac{1}{3}\overline{\sum_{i,j,k}B_{ij}B_{jk}B_{ki}} \nn\\
    =&-\frac{1}{3}\displaystyle\sum_{\substack{
        i=1;j=2;k=1,\ldots,N \\
        i=2;j=1;k=1,\ldots,N \\
        j=1;k=2;i\neq 1 \\
        j=2;k=1;i\neq 2 \\
        k=1;i=2;j\neq 1,2 \\
        k=2;i=1;j\neq 1,2 \\
      }}B_{ij}B_{jk}B_{ki} \label{eq:TE_m3_sum}\\
    =&-[(\Omega C)_{XY}]^{2}(\Omega_{YY}-1)-[(\Omega C)_{XY}]^{2}(\Omega_{XX}-1) \nn\\
    &-2[(\Omega C)_{XY}][(\Omega C)_{YY}]\Omega_{YX} \nn\\
    &-2[(\Omega C)_{XY}][(\Omega C^2)_{YY}][(\Omega C)_{YX}] \nn\\
    &-2\sum_{l>2}[(\Omega C)_{XY}][(\Omega C^l)_{YY}][(\Omega C^{l-1})_{YX}]. \label{eq:TE_m3_terms}
\end{align}
\end{subequations}
The six cases in the sum in~\eq{TE_m3_sum} are those where at least one of the indices~($i,j,k$) is equal to~\num{1} and another index is equal to~\num{2} (the third index can range between~\num{1} and~$N$, with some values excluded to avoid double counting).
%where the last step follows from symmetry of~$B$.
Plugging \eq{omega_series} into \eq{TE_m3_terms} yields
\begin{subequations}\label{eq:motifs_3}
    \begin{align}
        T_{X\rightarrow Y}^{\,(3)}
        =& - (C_{XY})^{2} (C^{T}C)_{XX} -(C_{XY})^{2} (C^{T}C)_{YY} \nn\\
        &-2 C_{XY} C_{YY} (C^{T}C)_{YX} - 2 C_{XY} (C^2)_{YY} C_{YX} \nn\\
        &+\mathcal{O}(\|C\|^6) \nn\\
        =&-\sum_{i_1} (C_{XY})^2 (C_{i_1,X})^2 \label{eq:motifs_3_1}\\
        &-\sum_{i_1} (C_{XY})^2 (C_{i_1,Y})^2 \label{eq:motifs_3_2}\\
        &-2 \sum_{i_1} C_{XY} C_{YY} C_{i_1X} C_{i_1Y} \label{eq:motifs_3_3}\\
        &-2 \sum_{i_1} C_{XY} C_{YX} C_{Yi_1} C_{i_1Y} \label{eq:motifs_3_4}\\
        &+\mathcal{O}(\|C\|^6). \nn
    \end{align}
\end{subequations}
Similarly, when $m=4$, we have
\begin{subequations}\label{eq:motifs_4}
    \begin{align}
        T_{X\rightarrow Y}^{\,(4)}=&\frac{1}{4}\overline{\textnormal{tr}[B^4]}=\frac{1}{4}\overline{\sum_{i,j,k,l}B_{ij}B_{jk}B_{kl}B_{li}} \nn\\
        =&\frac{1}{2}(C_{XY})^{4} \label{eq:motifs_4_1}\\
        &+ (C_{XY})^{2} (C_{YY})^{2} \label{eq:motifs_4_2}\\
        &+ (C_{XY})^{2} (C_{YX})^{2} \label{eq:motifs_4_3}\\
        &+2 C_{XY} C_{YX} (C_{YY})^{2} \label{eq:motifs_4_4}\\
        &+\mathcal{O}(\|C\|^6). \nn
    \end{align}
\end{subequations}
The full derivation for the case $m=4$ is provided in \appRef{m4}. We will not need to consider the cases where $m>4$ since $T_{X\rightarrow Y}^{\,(m)}\in \mathcal{O}(\|C\|^6)\ \forall m>4$.

So far, we have analysed the cases $m=1,2,3,4$ separately. Let us now combine the results by summing the weighted walk counts from \Cref{eq:motifs_2,eq:motifs_3,eq:motifs_4}. In order to simplify the expressions, we will isolate the occurrences where the indices in the sums are equal to~$X$ or~$Y$
%(\ie where $i_1,i_2 \in \{X,Y\}$)
from the other values. In so doing, some of the weighted walk counts found previously will cancel each other. The final decomposition for the TE in terms of weighted walk counts of relevant motifs, which is the main result of this paper, is then
\begin{subequations}\label{eq:motifs_all}
    \begin{align}
        T_{X\rightarrow Y}=&\frac{1}{2}(T_{X\rightarrow Y}^{\,(2)}+T_{X\rightarrow Y}^{\,(3)}+T_{X\rightarrow Y}^{\,(4)})+\mathcal{O}(\|C\|^6) \nn\\
        =& +\frac{1}{2} (C_{XY})^2 - \frac{1}{4} (C_{XY})^4 \label{eq:motifs_all_a} \\
        & + \sum_{\substack{i_1 \neq X,Y \\ i_2 \neq X,Y,i_1}} C_{XY} C_{i_1X} C_{i_1i_2} C_{i_2Y} \label{eq:motifs_all_b} \\
        & +\frac{1}{2}\sum_{i_1 \neq X,Y} (C_{XY})^2 (C_{i_1X})^2 \label{eq:motifs_all_c} \\
        & -\frac{1}{2}\sum_{i_1 \neq X,Y} (C_{XY})^2 (C_{i_1Y})^2 \label{eq:motifs_all_d} \\
        & +\frac{1}{2} (C_{XX})^2 (C_{XY})^2 \label{eq:motifs_all_e} \\
        & + \sum_{i_1 \neq X,Y} C_{XY} C_{i_1i_1} C_{i_1X} C_{i_1Y} \label{eq:motifs_all_f} \\
        & + \sum_{i_1 \neq X,Y} C_{XY} C_{XX} C_{Xi_1} C_{i_1Y} \label{eq:motifs_all_g} \\
        & +\mathcal{O}(\|C\|^6). \nn
    \end{align}
\end{subequations}
The motifs from \Crefrange{eq:motifs_3_3}{eq:motifs_3_4} and \Crefrange{eq:motifs_4_2}{eq:motifs_4_4} were cancelled; on the other hand, the new motifs in \Crefrange{eq:motifs_all_e}{eq:motifs_all_g} were introduced as special cases of \eq{motifs_2_2}.
\eq{motifs_all_a} and \eq{motifs_all_d} are the only terms remaining from $T_{X\rightarrow Y}^{\,(3)}$ that are negatively correlated to TE and were not completely cancelled here.
\fig{motifs_all} provides a visual summary of the motifs involved in $T_{X\rightarrow Y}$, up to order $\mathcal{O}(\|C\|^4)$.

\section{Numerical simulations and discussion\label{sec:simulations}}

\subsection{Directed link}
The pairwise TE $T_{X\rightarrow Y}$ clearly depends on the weight of the directed link $X\rightarrow Y$ [as per \Cref{eq:motifs_all_a,eq:motifs_all_e} and corresponding \fig{motifs_all} (a, e)]. \eq{motifs_all_a} is the dominant term in \eq{motifs_all} for linear Gaussian systems with weights $C_{XY}\in [-1,1]$ being similar across the network, which is perhaps not so surprising. %, being the only one of order $\mathcal{O}(\|C\|^2)$.
For such weights, the $(C_{XY})^2$ term will have a larger magnitude than the $(C_{XY})^4$ term, and so the total direct contribution of $C_{XY}$ to the TE in \eq{motifs_all_a} will be positive and increase with the magnitude of $C_{XY}$.
\subsubsection{Discussion}
Similarly, Hahs and Pethel~\cite{Hahs2013} analytically investigated the TE between coupled Gaussian processes -- for pairs of processes without a network embedding -- and identified a general increase with link weight.
Furthermore, a recent analytic study of a Boolean network model of policy diffusion also found that the TE depends on the square of the directed link weight as a first-order approximation~\cite{Goodman2020}.
Moreover, the directed link weight in the structural brain connectome is correlated with functional connectivity~\cite{Bettinardi2017,Honey2009}.
Positive or negative directed link weights result in the same contribution for the motifs in \Cref{eq:motifs_all_a,eq:motifs_all_e} (this dependence becomes more complex for higher order terms, see later sections). To distinguish the sign of the underlying link weight, one could examine the sub-components of the transfer entropy~\cite{Goetze2019}.

Yet, it is not always the case that information transfer is dominated by (or even correlated with) the weight of a directed link between the source and the target:
the dependence on the link weight is generally non-monotonic, especially in nonlinear systems (see~\cite{Schreiber2000} and~\cite[Fig 4.1]{Bossomaier2016}).

\subsection{In-degree of source and target\label{sec:in-degrees}}
Beyond the effect of the directed link, the TE increases with the in-degree of the source~$X$ [see \eq{motifs_all_c} and \fig{motifs_all}(c)] and decreases with the in-degree of the target~$Y$ [see \eq{motifs_all_d} and \fig{motifs_all}(d)], regardless of the sign of the weights (since the weights are squared in the sums).
This is because a higher number of incoming links can increase the variability of the source $X$ (and therefore its entropy), which enables higher TE. The same effect has the opposite consequence on the target: although a higher target in-degree may increase the collective transfer~\cite{Lizier2010,Lizier2008} from the set of sources taken jointly, the confounds introduced by more sources weaken the predictive effect of each single source considered individually.
The result is an asymmetry of information transfer, whereby the TE from the hubs to the other nodes is larger than the TE from the other nodes to the hubs. These factors are expected to have a strong effect in networks with low clustering coefficient, where the other motifs [\Cref{eq:motifs_all_b,eq:motifs_all_f,eq:motifs_all_g}] are comparatively rare on average, \eg in random networks.

\subsubsection{Numerical simulations\label{sec:in-degrees_numerical}}

In order to test this prediction, the TE between all pairs of linked nodes was measured in undirected scale-free networks of \num{100} nodes obtained via preferential attachment~\cite{Barabasi1999}. At each iteration of the preferential attachment algorithm, a new node was connected bidirectionally to a single existing node (as well as to itself via a self-loop). A constant uniform link weight $C_{XY}=C_{XX}=0.1$ was assigned to all the links, including the self-loops. The theoretical TE was computed according to \eq{TE_logs} with $k=14$ (matching the later empirical studies in \secRef{clustered_motifs}) and approximating $\Omega$ via the power series in \eq{omega_series} (until convergence). Differently from~\eq{motifs_all}, the higher order terms (\ie $\mathcal{O}(\|C\|^6)$) are not neglected. The experiment was repeated on \num{10000} different realisations of scale-free networks and the TE was averaged over the pairs with the same source and target in-degrees.
\begin{figure}
    \ifarXiv\includegraphics[width=0.48\textwidth]{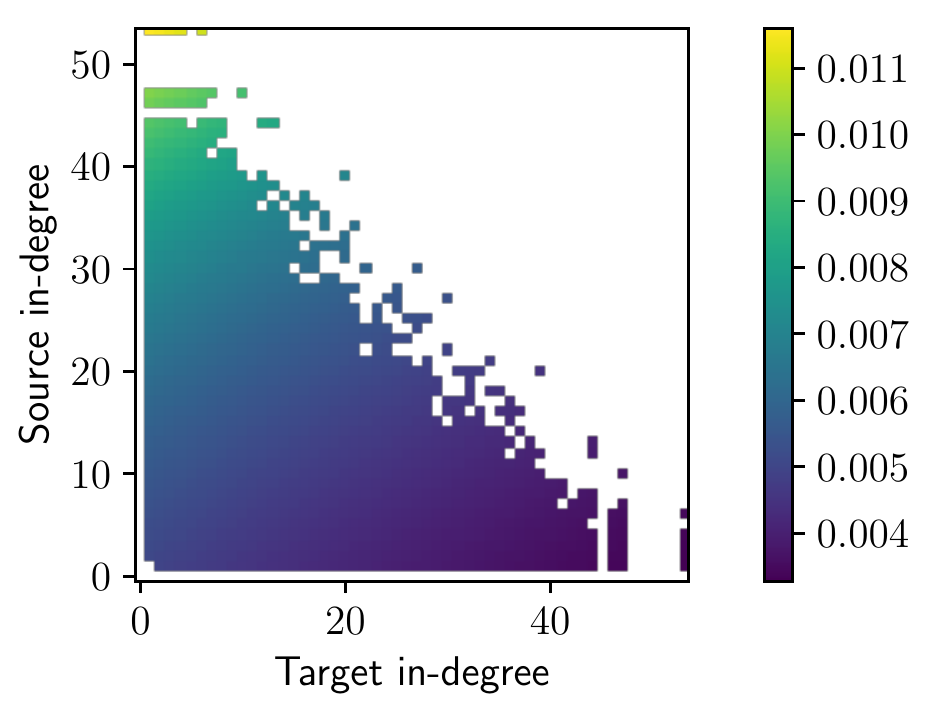}
    \else\centering\includegraphics[width=0.8\textwidth]{BA_bTE_vs_indegrees}
    \fi
    \caption{\label{fig:BA_bTE_vs_indegrees}
        The pairwise transfer entropy (TE) increases with the in-degree of the source and decreases with the in-degree of the target, regardless of the sign of the link weights.
        The TE is plotted as a function of the source and target in-degree. The results were obtained from \num{10000} simulations of scale-free networks of~\num{100} nodes generated via preferential attachment and the TE was averaged over all the node pairs with the same source and target in-degree. Note that the values in the lower left corner are the result of an average over many samples, since most of the node pairs have low in-degree. There are progressively fewer samples for higher in-degree pairs, and none for most pairs in the upper-right corner (absence indicated by the white colour).
    }
\end{figure}

% Discuss scale-free results
As shown in \fig{BA_bTE_vs_indegrees}, the pairwise TE increased with the source in-degree and decreases with the target in-degree. The factor-of-three difference between the minimum and maximum TE values underlines the importance of these network effects beyond local pairwise link weights.

\subsubsection{Discussion}

Interestingly, qualitatively similar results were obtained when the experiment was replicated on random Boolean networks, despite their nonlinear dynamics (\appRef{RBN}).
Similarly, a recent analytic study of a Boolean network model of policy diffusion also found that the TE is proportional to the weighted in-degree of the source and negatively proportional to the weighted in-degree of the target, as a second-order approximation~\cite{Goodman2020}.
A positive correlation between the pairwise TE and the in-degree of the source was also reported in simulations involving neural mass models~\cite{Li2019}, Kuramoto oscillators~\cite{Ceguerra2011}, and a model of cascading failures in energy networks~\cite{Lizier2009}. This is consistent with further findings showing that the degree of a node~$X$ is correlated to the ratio of (average) outgoing to incoming information transfer from/to~$X$ in various dynamical models, including Ising dynamics on the human connectome~\cite{Marinazzo2014,Marinazzo2012}.
Similarly, a study by Walker et al.~\cite{Walker2016} on effects of degree-preserving versus non-degree-preserving network randomisations on Boolean dynamics suggests that the presence of hubs plays a significant role in information transfer, as well as identifying that local structure beyond degree also contributes (as per the next section). 
Our results reinforce the suggestion that such correlation of source in-degree to TE is to be expected in general~\cite{Li2019}, since the linear Gaussian autoregressive processes considered here can be seen as approximations of nonlinear dynamics in the weakly coupled near-linear regime~\cite{Barnett2009a}. 
%and also because sources with higher in-degree have greater diversity of inputs and so potentially more predictive power.

Differently though, Timme et al.~\cite{Timme2016} report that the out-degree of the source correlates with the computation performed by a neuron (defined as the synergistic component of the TE \cite{Williams2011}).
It is difficult to interpret a direct mechanistic reason for this, however it is possible that this effect is mediated indirectly by re-entrant walks between the source and the target, similarly to how the path-transitivity enhances the undirected functional connectivity~\cite{Goni2014}. 
The role of the motifs involving multiple walks is discussed in the next section. 

Returning to the earlier qualification that a higher target in-degree may increase the collective transfer from the target's set of sources taken jointly, we note that this was previously empirically observed by Li et al.~\cite{Li2019}, and over the sum of pairwise transfers by Olin-Ammentorp and Cady~\cite{Olin-Ammentorp2018}.
Analytically investigating collective transfer across a set of sources jointly for the VAR dynamics remains a topic for future work.

Finally, echoing~\cite{Goodman2020}, the effect of the in-degree has implications for computing the directed functional connectivity via the pairwise TE, which has been widely employed in neuroscience~\cite{Honey2007,Ito2011,Stetter2012,Wibral2014}. When using TE as a pairwise measure, the links from hubs to low-degree nodes would generally be easier to infer than links between hubs, as well as links from low-degree nodes to hubs. This applies especially when the low number of time samples makes it difficult to distinguish weak transfer from noise and, importantly, could introduce a bias in the estimation of network properties.
More specifically, we expect the in-degree of hubs to be underestimated, which may thin the tail of the in-degree distribution. As Goodman and Porfiri~\cite{Goodman2020} also concluded, ``the out-degree plays a surprisingly marginal role on the quality of the inference". However, where the out-degree is correlated to the in-degree (\eg for undirected networks), we expect the out-degree of non-hubs to be underestimated, which may relatively fatten the tail of the out-degree distribution. 
For all of these reasons, the rich-club coefficient~\cite{VandenHeuvel2011} may also be altered.
These implications also apply to iterative or greedy algorithms based on multivariate TE~\cite{Faes2011,Lizier2012Multivariate,Montalto2014,Sun2015,Novelli2019}, since they rely on computing the pairwise TE as a first step.

\subsection{Clustered motifs\label{sec:clustered_motifs}}
So far, we have discussed the directed motif [\eq{motifs_all_a}] and we have considered networks with low global clustering coefficient, where the in-degree of the source and the target [\Cref{eq:motifs_all_c,eq:motifs_all_d}] play an important role. In networks with higher global clustering coefficients, such as lattice or small-world networks, other motifs will provide a significant contribution to the pairwise TE beyond the effect of the in-degrees. Specifically, these are the \emph{clustered} motifs that involve a common-parent [\Cref{eq:motifs_all_b,eq:motifs_all_f} and corresponding \fig{motifs_all} (b, f)] or a secondary path [\eq{motifs_all_g} and \fig{motifs_all}(g)] in addition to the directed link $X\rightarrow Y$.
The relative importance of the terms in \eq{motifs_all} depends in fact on the properties of the network: if the clustering coefficient is high, the abundance of the clustered motifs makes their effect significant, despite each motif only contributing to the TE at order~\num{4} [see \Cref{eq:motifs_all_b,eq:motifs_all_f,eq:motifs_all_g}].
Therefore, if the link weights are positive, we would expect the pairwise TE to be higher (due to these motifs) than what would be accounted for by the directed and in-degree motifs alone.
The reason is that the common parent and the secondary pathways \textit{reinforce} the effect of the directed link $X\rightarrow Y$, leading to a greater predictive payoff from knowing the activity of the source $X$.

\subsubsection{Numerical simulations\label{sec:clustered_motifs_numerical}}

\begin{figure}
    \ifarXiv\includegraphics[width=0.48\textwidth]{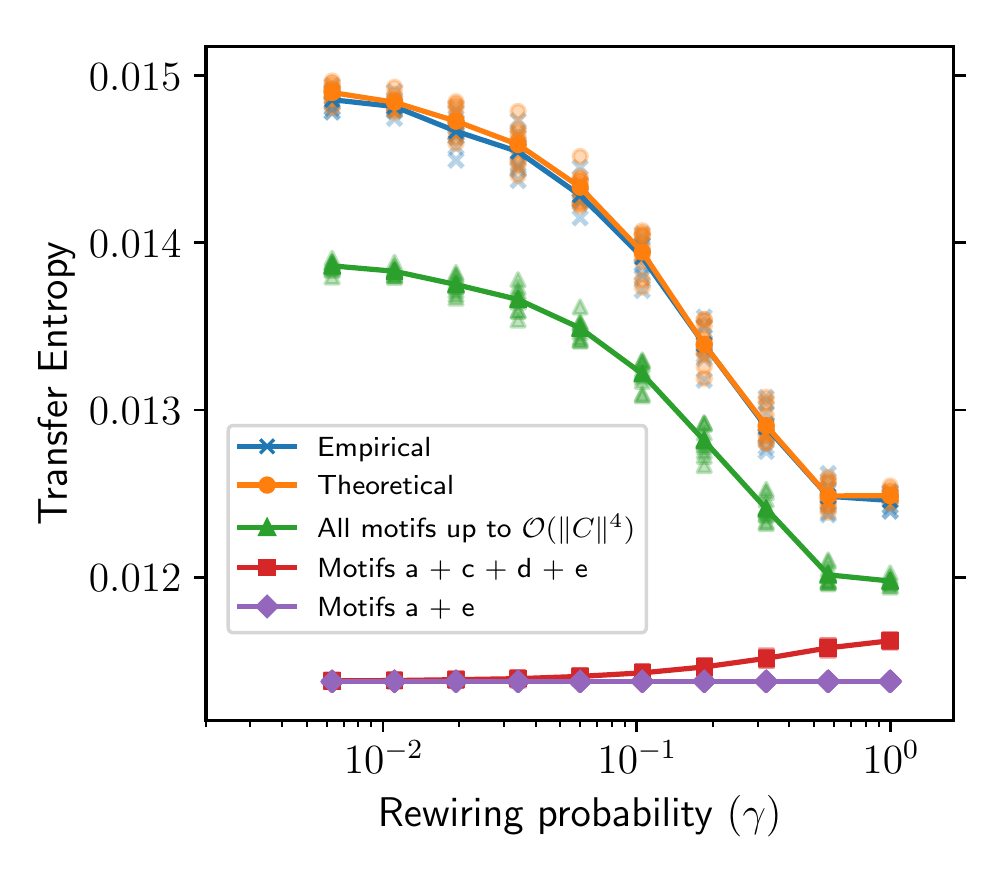}
    \else\centering\includegraphics[width=0.8\textwidth]{WS_bTE_vs_rewiring_w015}
    \fi
    \caption{\label{fig:WS_bTE_vs_rewiring}
        Average transfer entropy (TE) as a function of the rewiring probability in Watts-Strogatz ring networks.
        For positive link weights, the pairwise TE is higher in clustered networks than in random networks, due to the higher number of clustered motifs.
        For each value of the rewiring probability ($\gamma$), the results for~\num{10} simulations on different networks are presented (low-opacity markers) in addition to the mean values (solid markers). The plot shows that the approximation based on all the motifs up to order~\num{4} (green curve) is closer to the theoretical values (orange curve) than the approximation based on the in-degrees and directed motifs alone (red curve) or on the directed motifs alone (violet curve). The empirical values are also shown (blue curve) as a validation of the theoretical results.
    }
\end{figure}

This prediction was tested on Watts-Strogatz ring networks~\cite{Watts1998}, starting from a directed ring network of $N=100$ nodes with uniform link weights $C_{XY}=C_{XX}=0.15$ and fixed in-degree~$d_\textnormal{in}=4$ (\ie each node was linked to two neighbours on each side as well as itself). The source of each link was rewired with probability~$\gamma$, such that the in-degree of each node was unchanged and the effect of the other motifs could be studied. 
The clustering coefficient decreased for higher values of~$\gamma$ as the network underwent a small-world transition, and so did the number of clustered motifs. Accordingly, the average theoretical TE between linked nodes (computed via~\eq{TE_logs} with $k=14$ as above) decreased as predicted (see orange curve in \fig{WS_bTE_vs_rewiring}).

\fig{WS_bTE_vs_rewiring} also reports the empirical values of the TE, estimated from synthetic time series of \num{100000} time samples. The analysis was carried out using the IDTxl software~\cite{Wollstadt2019}, employing the Gaussian estimator and selecting an optimal embedding of size~$k=14$ for the target time series~\footnote{The determination of these embedding parameters followed the method of Garland et al.~\cite{Garland2016} finding the values which maximise the active information storage, with the important additional inclusion of bias correction (because increasing $k$ generally serves to increase bias of the estimate)~\cite{Erten2017}.}. This provides a validation of the theoretical TE [computed via~\Cref{eq:TE_logs,eq:omega_series}], which matches these empirical values. The approximation in terms of motifs up to order $\mathcal{O}(\|C\|^4)$ [computed via~\eq{motifs_all}], while not capturing all higher order components of the TE, do reproduce the overall trend in agreement with the theoretical values, providing further validation of our main derivations. On the other hand, the partial approximation based on the directed link weight and the in-degree (motifs a, c, d, and e) is not sufficient to reproduce the empirical TE trend, since that partial approximation does not account for the changing contribution of motif structures with the rewiring parameter $\gamma$.

\subsubsection{Discussion}

If the link weights are positive, the pairwise TE increases with the \textit{number} of clustered motifs. (This applies on average in the mammalian cortex, where the majority of the connections are thought to be excitatory~\cite{Barnett2011}.) As such, the effect of the clustered motifs has implications for computing the directed functional connectivity via the pairwise TE: the directed functional connectivity is better able to infer links within brain modules (where such motifs enhance TE values) than links across modules.
This appears to align with results of Stetter et al.~\cite{Stetter2012}, finding that the true positive rate for TE based directed functional network inference on simulated neural cultures generally increased with clustering coefficient of the underlying network structure.
When negative weights are present (interpretable as inhibitory in a neural context), the direct relationship to the \textit{number} of motifs for \Cref{eq:motifs_all_b,eq:motifs_all_f,eq:motifs_all_g} is less clear and depends intricately on the proportion and placement of these negatively-weighted links (though the overall relation to weighted motif counts obviously still holds).

Differently from the case of the in-degree, the effect of the clustered motifs on the pairwise TE was not qualitatively preserved in random Boolean networks.
Our experiments on RBNs in \appRef{RBN} show that the pairwise TE increases with the rewiring probability~$\gamma$ there.
These results align with more comprehensive experiments in a previous study~\cite{Lizier2011}. There, it was argued that long links are able to introduce new information to the target that it was less likely to have previously been exposed to, in contrast to information available from its clustered near neighbours. This effect does not appear to be so important for linear dynamics, as it cannot be identified in the motifs in \eq{motifs_all} and \fig{motifs_all}.
Mediano and Shanahan~\cite{Mediano2017} also report a slightly different effect in other non-linear dynamics. That is, that averages of (higher-order conditional) TE peaks at values of $\gamma$ on the random side of the small-world regime in a model of coupled spiking neurons (in contrast to our approach, this is averaged over all pairs of nodes in the system, connected or not). They argue that the neurons are functionally decoupled in the regular regime, and that in the random regime the strong correlations across the network mean that the source cannot add information about the target beyond what is already conditioned on.
The dominant effect in the linear dynamics under consideration here are the reinforcements achieved from clustered structure identified in \Cref{eq:motifs_all_b,eq:motifs_all_f} and \eq{motifs_all_g}; that is an \textit{additive} reinforcement effect, and so is likely less pertinent to non-linear dynamics such as in RBNs and spiking neurons.

\subsection{Further remarks}
The decomposition of the pairwise TE in terms of network motifs [\eq{motifs_all} and \fig{motifs_all}] was performed up to order $\mathcal{O}(\|C\|^4)$. Longer motifs will start to appear in higher-order approximations. For example, motifs involving a confounding effect (\ie a common parent of $X$ and $Y$ without the directed link $X\rightarrow Y$) appear at order \num{6} (not shown). The higher order motifs are providing only a small contribution for $C_{XY}=C_{XX}=0.15$ in \fig{WS_bTE_vs_rewiring}; that contribution will become more significant as link weights become larger (in particular when the spectral radius is close to~\num{1}).

A similar decomposition of the the active information storage in the dynamics of a target node was provided in previous work~\cite{Lizier2012}, reporting that the highest order contributions were from low-order feedback and feed-forward motifs (with the relevant feed-forward motifs converging on the target node $Y$).
The motifs contributing to the information storage at a node $Y$ contrast to those contributing to the decomposition of information transfer from $X \rightarrow Y$ presented in~\eq{motifs_all}. First, there is no explicit contribution of feedback loops in the TE decomposition. This may seem contrary to the expectation of their \textit{detracting} from TE (since they facilitate prior knowledge of the source stored in the past of the target, which TE removes). While such terms do not appear explicitly, their detracting effect has been implicitly removed prior to the final result: because the unlabelled nodes in \fig{motifs_all} are distinct from the target $Y$, any feedback loops potentially including $Y$ have been removed from the counts in \fig{motifs_all} (panels b, f, g).
Moreover, the types of feed-forward motifs that contribute to information storage on $Y$ and transfer from $X \rightarrow Y$ are slightly distinct. Feed-forward motifs contribute to transfer here where the source $X$ is on one of two walks with the \textit{same lengths} to $Y$ from some common driver [\Cref{eq:motifs_all_b,eq:motifs_all_f,eq:motifs_all_g}]. In contrast, a motif will generate an information storage effect on the target $Y$ where the lengths of those walks are distinct~\cite{Lizier2012}. We can interpret this as the difference between the reinforcement of a direct effect from $X$ (transfer) versus a correlation in $Y$ of dynamics across time steps (storage).

% dependence of info-theoretic measures on global structure

%on non-local effects: Remotely induced rerouting of information in modular networks. "local interventions within one sub-network may remotely determine nonlocal network-wide communication"~\cite{Kirst2016}

\section{Conclusion}
A linear, order-\num{1} autoregressive process was used to systematically investigate the dependence of the pairwise transfer entropy (TE) on the global network topology. 
%beyond the local effect of the link weights.
Specific weighted motifs were found to enhance or reduce the TE [\eq{motifs_all}], as summarised in \fig{motifs_all}.
The assumptions of linearity, stationarity, Gaussian noise, and uniform link weights were made in order to enable the analytical treatment. Importantly, under these assumptions, the results also apply to Granger causality~\cite{Barnett2009b}. Moreover, the numerical simulations in \appRef{RBN} and the recent literature on the topic suggest that the dependence of the TE on the in-degree also holds for nonlinear dynamics.

In future work, the analytic approach will be extended to linear systems in continuous time, such as the multivariate Ornstein-Uhlenbeck process (as performed by Barnett et al.~\cite{Barnett2009a,Barnett2011} for the Tononi-Sporns-Edelman (TSE) complexity~\cite{Tononi1994}). Recent progress has already been made in the inference of the weighted adjacency matrix from observations for these continuous-time systems~\cite{Ching2015,Ching2017,Zhang2015}.
Furthermore, higher order conditional and collective transfer entropies~\cite{Lizier2008,Lizier2010} could also be investigated in a similar fashion. Since conditional TE terms remove redundancies and include synergies between the considered source and conditional sources~\cite{Williams2011}, it is likely that there will be both removal of previous and inclusion of new contributing motif structures in comparison to the pairwise effect.

\appendix
\ifarXiv
\else
\section*{Appendices}
\fi

\section{Covariance matrices and non-zero terms in \eq{TE_traces}}
\label{app:cov_matrices}

The covariance matrices $\Omega(Y,\bm{Y^{(k)}})-I,\ \Omega(\bm{Y^{(k)}})-I,\text{ and }\Omega(X,\bm{Y^{(k)}})-I$ can be obtained as submatrices of $B=\Omega(X,Y,\bm{Y^{(k)}})-I$ [see \eq{B}]. Specifically, we have:
\begin{equation}
\Omega(\bm{Y^{(k)}})-I=
    \left(
    \begin{array}{ccc}
        \Omega(0)_{YY}-1 & \cdots & \Omega(k-1)_{YY}\\
        \vdots  & \ddots & \vdots \\
        \Omega(k-1)_{YY} & \cdots & \Omega(0)_{YY}-1
    \end{array}
    \right)
\end{equation}

\begin{multline}
\Omega(Y,\bm{Y^{(k)}})-I=\\
    \left(
    \begin{array}{c|ccc}
        \Omega(0)_{YY}-1 & \Omega(1)_{YY} & \cdots & \Omega(k)_{YY} \\
        \hline
        \Omega(1)_{YY} & \Omega(0)_{YY}-1 & \cdots & \Omega(k-1)_{YY} \\
        \vdots  & \vdots & \ddots & \vdots \\
        \Omega(k)_{YY} & \Omega(k-1)_{YY} & \cdots & \Omega(0)_{YY}-1
    \end{array}
    \right)
\end{multline}

\begin{multline}
\Omega(X,\bm{Y^{(k)}})-I=\\
    \left(
    \begin{array}{c|ccc}
        \Omega(0)_{XX}-1 & \Omega(0)_{YX} & \cdots & \Omega(k-1)_{YX} \\
        \hline
        \Omega(0)_{YX} & \Omega(0)_{YY}-1 & \cdots & \Omega(k-1)_{YY} \\
        \vdots & \vdots  & \ddots & \vdots \\
        \Omega(k-1)_{YX} & \Omega(k-1)_{YY} & \cdots & \Omega(0)_{YY}-1
    \end{array}
    \right)
\end{multline}
The four matrix traces involved in \eq{TE_traces} are
\begin{subequations}
\begin{alignat}{2}
    &\textnormal{tr}[(\Omega(Y,\bm{Y^{(k)}})-I)^m], \label{eq:tr_a}\\
    &\textnormal{tr}[(\Omega(\bm{Y^{(k)}})-I)^m], \label{eq:tr_b}\\
    &\textnormal{tr}[(\Omega(X,Y,\bm{Y^{(k)}})-I)^m]=\textnormal{tr}[B^m], \label{eq:tr_c}\\
    &\textnormal{tr}[(\Omega(X,\bm{Y^{(k)}})-I)^m]. \label{eq:tr_d}
\end{alignat}
\end{subequations}
Let us start with the difference [\eq{tr_d} - \eq{tr_c}]. The trace in \eq{tr_c} can be expanded as
\begin{align}\label{eq:trace_as_sum}
    \textnormal{tr}[B^m]=&\sum_{i}(B^m)_{ii} \nn\\
    %=&\sum_{i_1,\ldots,i_{m}=1}^N B_{i_{1}}B_{i_{2}}\left(\prod_{j=2}^{m-1}B_{i_{j}i_{j+1}}\right)B_{i_{m}}B_{i_{1}} \nn\\
    =&\sum_{i_1,\ldots,i_{m}} B_{i_{1}i_{2}}B_{i_{2}i_{3}}\ldots B_{i_{m-1}i_{m}}B_{i_{m}i_{1}}
\end{align}
and the trace in \eq{tr_d} can be expanded similarly as a sum. With $\Omega(X,\bm{Y^{(k)}})-I$ being a submatrix of~$B$, all the terms in \eq{tr_d} also appear in \eq{tr_c}. Thus, the remaining terms in the difference [\eq{tr_d} - \eq{tr_c}] are the terms in \eq{trace_as_sum} that involve entries from the second row (or column) of~$B$, \ie those where at least one of the indices~$i_1,\ldots,i_{m}$ is equal to~$2$ (corresponding to $Y$).

Similarly, all the terms in \eq{tr_b} also appear in \eq{tr_a}. Thus, the remaining terms in the difference [\eq{tr_a} - \eq{tr_b}] are those where at least one of the indices~$i_1,\ldots,i_{m}$ corresponds to $Y$ (being equal to~$1$ for the matrix in \eq{tr_a}, but equal to~$2$ when aligned with matrix $B$ in \eq{trace_as_sum}).

Finally, the remaining terms in the trace differences in \eq{TE_traces}
$$
\textnormal{[\eq{tr_a} - \eq{tr_b}] - [\eq{tr_c} - \eq{tr_d}]}
$$
are the terms in \eq{trace_as_sum} that i. involve at least one entry of~$B$ from the second row (or column) corresponding to $Y$ (as per the arguments above), and also ii. involve at least one entry of~$B$ from the first row (or column) corresponding to $X$ (in order to appear in $\textnormal{[\eq{tr_c} - \eq{tr_d}]}$ but not $\textnormal{[\eq{tr_a} - \eq{tr_b}]}$).
That is, the remaining terms are those in \eq{trace_as_sum} where at least one of the indices~$i_1,\ldots,i_{m}$ is equal to~\num{1} and another one is equal to~$2$.

\section{Derivation of motifs for $m=4$}
\label{app:m4}
When $m=4$ in~\eq{TE_B}, we have
\begin{align}
    T_{X\rightarrow Y}^{\,(4)}=&\frac{1}{4}\overline{\textnormal{tr}[B^4]}=\frac{1}{4}\overline{\sum_{i,j,k,l}B_{ij}B_{jk}B_{kl}B_{li}},
\end{align}
where the overbar indicates that only the terms that involve at least one entry of~$B$ from the first row and one from the second row (or columns) are considered. There are~\num{12} cases to consider, \ie those where at least one of the four indices~($i,j,k,l$) is equal to~\num{1} and another index is equal to~\num{2} (the other indices can range between~\num{1} and $N$, with some values excluded to avoid double counting):
%Due to symmetry of~$B$, all of the~\num{12} cases can be rewritten in one of these two forms:
%\begin{subequations}
%    \begin{align}
%        \sum_{k,l}B_{12}B_{2k}B_{kl}B_{l1} %\label{eq:motifs_4_form_1}\\
%        \sum_{k,l}B_{1k}B_{k2}B_{2l}B_{l1} %\label{eq:motifs_4_form_2}
%    \end{align}
%\end{subequations}
%The four cases which are equivalent to \eq{motifs_4_form_2} can be neglected since they contribute at order $\mathcal{O}(\|C\|^6)$. We are then left with the~\num{8} cases in the form \eq{motifs_4_form_1}:
\begin{subequations}
    \begin{align}
        T_{X\rightarrow Y}^{\,(4)}=&\frac{1}{4}\displaystyle\sum_{\substack{
            i=1;j=2;k;l \\
            i=2;j=1;k;l \\
            i\neq 2;j=1;k=2;l \\
            i\neq 1;j=2;k=1;l \\
            i;j\neq 1,2;k=1;l=2 \\
            i;j\neq 1,2;k=2;l=1 \\
            i=2;j\neq 1;k\neq 1,2;l=1 \\
            i=1;j\neq 2;k\neq 1,2;l=2 \\
        }}B_{ij}B_{jk}B_{kl}B_{li} \label{eq:motifs_4_form_1} \\
        &+\frac{1}{4}\displaystyle\sum_{\substack{
            i=1;k=2;j;l \\
            i=2;k=1;j;l \\
            j=1;l=2;i\neq 1;k\neq 1 \\
            j=2;l=1;i\neq 2;k\neq 2 \\
        }}B_{ij}B_{jk}B_{kl}B_{li}. \label{eq:motifs_4_form_2}
    \end{align}
\end{subequations}
The terms in \eq{motifs_4_form_2} will be neglected since they contribute at order $\mathcal{O}(\|C\|^6)$ once the expansions of the covariance matrices are inserted [\Cref{eq:omega_series,eq:omega_lag}]. Computing the remaining terms in \eq{motifs_4_form_1} gives the result shown in \eq{motifs_4}.

\section{Extension to Random Boolean Networks}
\label{app:RBN}
Random Boolean Networks are a class of discrete dynamical systems which were proposed as models of gene regulatory networks by Kauffman~\cite{Kauffman1993}. Each node in the network has a Boolean state value, which is updated in discrete time. In the original formulation, the new state of each node is a deterministic Boolean function of the current state of its parents. Given the topology of the network, this function is assigned at random for each node when the network is initialised, subject to a probability~$r$ of producing ``\num{1}" outputs. Differently from the original formulation, the Boolean function was made stochastic here by introducing a probability~$p=0.005$ of switching state at each time step.

\begin{figure}[t]
    \ifarXiv\includegraphics[width=0.5\textwidth]{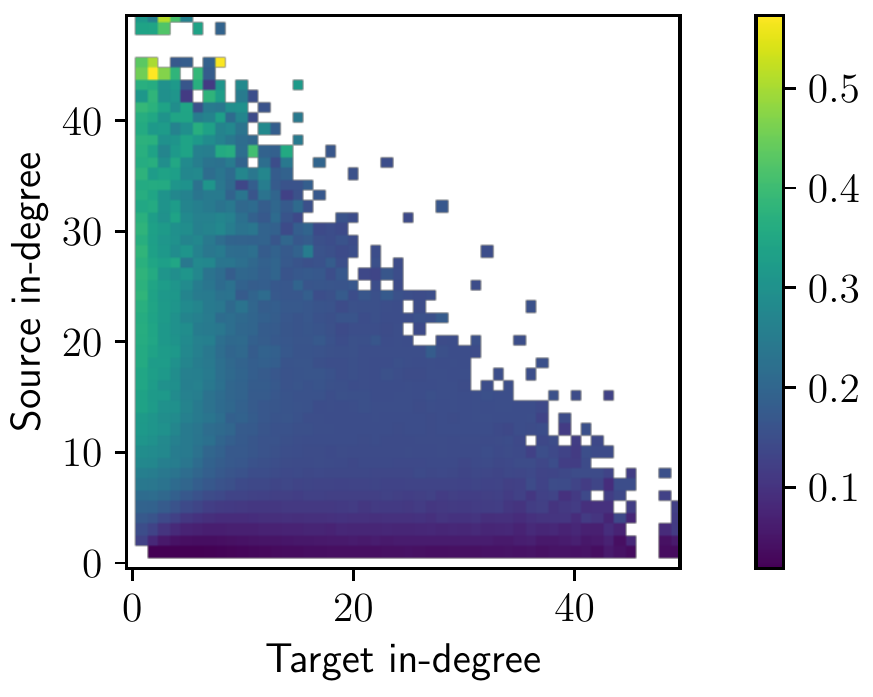}
    \else\centering\includegraphics[width=0.8\textwidth]{BA_bTE_vs_indegrees_boolean}
    \fi
    \caption{\label{fig:BA_bTE_vs_indegrees_boolean}
        Pairwise transfer entropy (TE) as a function of the source and target in-degrees in random Boolean networks.
        Similarly to the linear Gaussian case (\fig{BA_bTE_vs_indegrees}), the TE increases with the in-degree of the source and decreases with the in-degree of the target.
        The results were obtained from \num{10000} simulations of scale-free networks of~\num{100} nodes generated via preferential attachment. The TE was averaged over all the node pairs with the same in-degrees. The values in the lower left corner are the result of an average over many samples, since most of the node pairs have low in-degrees. There are progressively fewer observations for higher in-degrees and none in the upper-right corner (absence indicated by white colour).
    }
\end{figure}

The experiment described in \secRef{simulations} (In-degree of source and target) was repeated on Random Boolean Networks with~$r=0.5$ but keeping the same topology (scale-free networks obtained via preferential attachment). In the absence of theoretical results, the pairwise TE was estimated numerically from synthetic time series with \num{100000} time samples. The time series were embedded with a history length~$k=14$, as in \secRef{simulations}.
The results (shown in \fig{BA_bTE_vs_indegrees_boolean}) were qualitatively similar to those obtained using linear Gaussian processes (\fig{BA_bTE_vs_indegrees}).

\begin{figure}[t]
    \ifarXiv\includegraphics[width=0.48\textwidth]{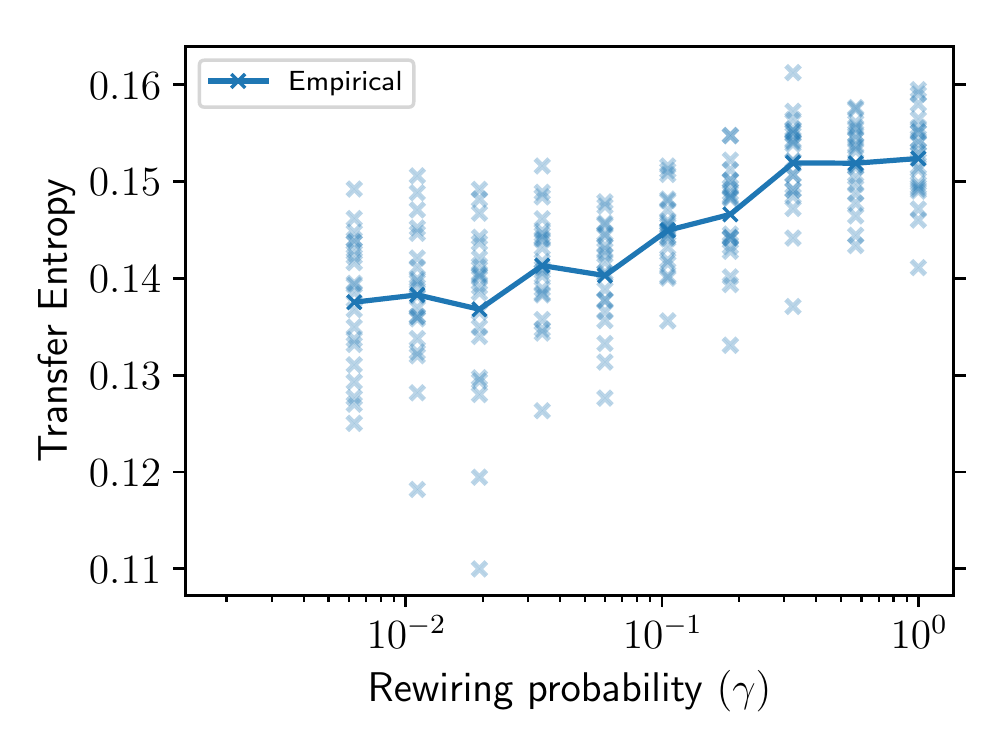}
    \else\centering\includegraphics[width=0.8\textwidth]{WS_bTE_vs_rewiring_boolean}
    \fi
    \caption{\label{fig:WS_bTE_vs_rewiring_boolean}
        Average empirical transfer entropy as a function of the rewiring probability in Watts-Strogatz ring networks with a random Boolean dynamics. The results for~\num{20} simulations on different networks are presented (low-opacity markers) in addition to the mean values (solid markers).
    }
\end{figure}

The experiment presented in \secRef{simulations} (Clustered motifs) was also repeated using the Random Boolean Networks but keeping the same topology. In this case, the results (shown in \fig{WS_bTE_vs_rewiring_boolean}) were not qualitatively similar to those obtained using linear Gaussian processes (\fig{WS_bTE_vs_rewiring}). As shown in previous studies~\cite{Lizier2011} (without the addition of stochastic noise), the pairwise TE increases with the rewiring probability~$\gamma$.

\ifarXiv
\begin{acknowledgments}
The authors acknowledge the Sydney Informatics Hub and the University of Sydney's high-performance computing cluster Artemis for providing the high-performance computing resources that have contributed to the research results reported within this paper.
\end{acknowledgments}
\else
\ack{
The authors acknowledge the Sydney Informatics Hub and the University of Sydney's high-performance computing cluster Artemis for providing the high-performance computing resources that have contributed to the research results reported within this paper.}
\fi

\ifarXiv
\section*{Funding statement}
JL was supported through the Australian Research Council DECRA Fellowship grant DE160100630 and through The University of Sydney Research Accelerator (SOAR) prize program.
\else
\funding{
JL was supported through the Australian Research Council DECRA Fellowship grant DE160100630 and through The University of Sydney Research Accelerator (SOAR) prize program.}
\fi

\ifarXiv
\section*{Author Contributions}
JL, LN, FA, and JJ conceptualised the study; LN and JL carried out the formal analysis; LN performed the numerical validation, prepared the visualisations, and wrote the original draft; all the authors edited and approved the final manuscript.
\else
\aucontribute{
JL, LN, FA, and JJ conceptualised the study; LN and JL carried out the formal analysis; LN performed the numerical validation, prepared the visualisations, and wrote the original draft; all the authors edited and approved the final manuscript.
}
\fi

\ifarXiv
\else
\section*{Data, code and materials}
The synthetic data was generated via computer simulations. The code and the data are available to facilitate the reproduction of the original results~\cite{Novelli2020ZenodoCode}.
\fi

\pagebreak

\ifarXiv
\else
\bibliographystyle{unsrt}
\fi
\bibliography{bibliography}

\end{document}